\documentclass{svjour3}                     
\smartqed  
\usepackage{graphicx}
\begin{document}

\title{Tsallis $q$-exponential describes the distribution of scientific citations - A new characterization of the impact}



\author{Aristoklis D. Anastasiadis, Marcelo P. de Albuquerque, Marcio P. de Albuquerque, Diogo B. Mussi 
}


\institute{A.D. Anastasiadis \at
Electrical and Computer Engineering Department, University of Patras, Rio Achaia, Greece, \\
Centro Brasileiro de Pesquisas Fisicas, Rua Xavier Sigaud 150 22290-180 Rio de Janeiro Brazil \\
              Tel.: +30-2610-997997 Fax: +30-2610-997997 \\
              \email{anastasiadis@ece.upatras.gr, anastasiadis@cbpf.br}
           \and
           Marcelo P. de Albuquerque \at
Centro Brasileiro de Pesquisas Fisicas, Rua Xavier Sigaud 150 22290-180 Rio de Janeiro Brazil \\
             \email{marcelo@cbpf.br}            \\
 \and
           Marcio P. de Albuquerque \at
Centro Brasileiro de Pesquisas Fisicas, Rua Xavier Sigaud 150 22290-180 Rio de Janeiro Brazil \\
             \email{mpa@cbpf.br}            \\
 \and
           Diogo B. Mussi \at
Centro Brasileiro de Pesquisas Fisicas, Rua Xavier Sigaud 150 22290-180 Rio de Janeiro Brazil \\
               \email{diogob@cbpf.br}            \\
}

\date{Received: date / Accepted: date}

\maketitle

\begin{abstract}
In this work we have studied the research activity for countries of Europe,
Latin America and Africa for all sciences between $1945$ and November $2008$. All the data are
captured from the Web of Science database during this period. The analysis of the experimental data shows that, within a nonextensive thermostatistical formalism, the Tsallis \emph{q}-exponential distribution $N(c)$ satisfactorily describes Institute of Scientific Information citations. The data which are examined in the present survey can be fitted successfully as a first approach by applying a {\it single} curve (namely, $N(c) \propto 1/[1+(q-1)\; c/T]^{\frac{1}{q-1}}$ with $q\simeq 4/3$ for {\it all} the available citations $c$,  $T$ being an ``effective temperature''. The present analysis ultimately suggests that the phenomenon might essentially be {\it one and the same} along the {\it entire} range of the citation number.
Finally, this manuscript provides a new ranking index, via the ``effective temperature'' $T$, for the impact level of the research activity in these countries,
taking into account the number of the publications and their citations.

\keywords{Citations; Nonextensive entropy; Web of Science; Complex systems} 
\end{abstract}

\section{Introduction}
\label{intro}
The analysis of the citations of scientific papers is an important issue that can enable a better understanding of the research activity
of the authors, the institutions and their countries~\cite{Hirsch05,Ball05,Times08,WOS}.
The evaluation of the productivity of individual scientists has traditionally relied on the number of papers they have published.
Nowadays, with the easy access to the Internet and to large databases,
including the Web of Science~\cite{WOS}, the comparison of the impact of scientific contributions is a much easier and more rapid process.

Many measures of research productivity have been proposed.
In many surveys, research productivity is usually represented by two different variables, namely
the number of total publications and their citations~\cite{LongCWD08}. The first measure reflects research quantity and the other reflects research impact. The degree to which published works are cited by other authors is generally considered as a reflection of the quality of those works~\cite{Phelan1999}.

There has been considerable work in the area of citation analysis. Prior citation
analysis has analyzed a wide variety of factors such as (i) the distribution of
citation rates \cite{Redner1998EPJB}\cite{LehmannLJ03},\cite{TsallisMA00},\cite{LaherrereS98}, (ii) the variation in the distribution of citation rates
across research fields and geographical regions \cite{LehmannLJ03}, (iii) the geographic distribution
of highly cited scientists \cite{Batty03a,Batty03b}, (iv) various indicators of the scientific
performance of countries \cite{May97}. Finally, citation analysis and other methodologies based on research productivity have been used
to rank journals~\cite{KaterattanakulHH03,Nwagwu06,SolariM02} and also universities~\cite{VogelW84,Shanghais08,Times08}.

The assessment of scientific research is an extremely delicate and sophisticated
venture~\cite{Braun1985}. The scientific position of a given country in the
international context can usually be analyzed from both qualitative and quantitative
points of view. Firstly, the number of publications of a country and its contribution to the
total world can be used. Secondly, the impact of its research outputs, preferably by
scientific disciplines, can be measured through citations or some other surrogate Impact
Factor measures~\cite{Hirsch05,WOS}.
Scientometric analysis plays an important role in the assessment of the performance
of scientific research, for it can address some structural problems such as the impact of research
outputs of some countries on several scientific fields, the scale and characteristics of the international
comparison, the structure of several fields, and the relationship which exists between
them~\cite{Hirsch05,WOS,BatistaCKM06,LongCWD08}.

With regard to the distribution of citations, many works have been done \cite{LaherrereS98,Redner1998EPJB,TsallisMA00}.
A stretched exponential fitting was applied for modeling citation distributions based on multiplicative processes
\cite{LaherrereS98}. Lehmann \cite{LehmannLJ03} attempted to fit both a power law and stretched exponential to the
citation distribution of 281\,717 papers in the SPIRES \cite{SpiresDatabase} database and showed
it is impossible to discriminate between the two models. Redner analyzed the ISI and Physical Review databases\cite{Redner1998EPJB}.
In Redner's work the applied fitting distribution had only partial success while the same numerical data for large citation count $c$ showed that can be fitted
quite satisfactorily with a single curve by using nonextensive thermostatistical formalism~\cite{TsallisMA00}.

In the present work, we have considered the scientific research activity in terms of the number of publications and number of citations. The current study uses data from Thomson ISI Web of Science database~\cite{WOS} for many different countries from Latin America, Europe and South Africa.  The period that is investigated is between 1945 to November 2008.
We show that the data for all the tested countries can be satisfactorily fitted with a {\it single} curve, namely $N(c) \propto 1/[1+(q-1)\; c/T]^{\frac{1}{q-1}}$ (with $q\simeq4/3$), which naturally emerges within the Tsallis theory.
The present analysis ultimately suggests  that this phenomenon might essentially be {\it one and the same}
along the {\it entire} range of the citation number $c$ for each different case.

\section{Nonextensive Statistical Mechanics and Web of Science Citations}
\label{Tsallis distribution}
Nowadays, the idea of nonextensivity has been used in many
applications. Nonextensive statistical mechanics has been applied successfully in physics (astrophysics, astronomy, cosmology,
nonlinear dynamics)~\cite{Shibata03,Siekman03}, biology~\cite{Upadhyaya01},
economics~\cite{Tsallis03}, human and computer sciences \cite{AnastasiadisM2004a,MarceloAM04,AnastasiadisM06,AlexandraTTMT04} and provide interesting insights into a variety of physical
systems (two-dimensional turbulence in pure-electron plasma \cite{boghosian}, variety of self-organized
critical models \cite{SOC}, long-range interaction conservative systems \cite{celia}, and among others \cite{biblio}).
Thomson ISI Web of Science \cite{WOS} is a widely used database source for such works.

\subsection{Nonextensive Statistical Mechanics}
Nonextensive statistical mechanics is based on Tsallis entropy. Tsallis statistics is currently considered useful in describing the thermostatistical properties of nonextensive systems; it is based on the generalized entropic form \cite{Tsallis88}:
\begin{equation}\label{eq:tsallis Sq}
    S_{q}\equiv k\; \frac{1-\sum_{i=1}^{W}p_{i}^{q}}{q-1} \;\;\;\;
(q \in \Re),
\end{equation}
where $W$ is the total number of microscopic configurations, whose probabilities are $\{p_i\}$, and $k$ is a conventional positive constant. When $q=1$ it reproduces the  Boltzmann-Gibbs entropic form $S_{BG}=-k\sum_{i=1}^W p_i \ln p_i$. The nonextensive entropy $S_{q}$ achieves its extreme value at the equiprobability $p_{i}=1/W,\forall i$, and this value equals $S_{q}=k \frac{W^{1-q}-1}{1-q}$ ($S_1=S_{BG}=k \ln W$)~\cite{Tsallis88,GellMannT04}.
The Tsallis entropy is nonadditive in such a way that, for statistical independent systems $A$ and $B$, the entropy satisfies the following property:
\begin{equation} \label{nonadditive}
\frac{S_q(A+B)}{k}=\frac{S_q(A)}{k} + \frac{S_q(B)}{k} + (1-q)\frac{S_q(A)}{k} \frac{S_q(B)}{k} \,.
\end{equation}
 It is subadditive for $q>1$, superadditive for $q<1$, and, for $q=1$, it recovers the BG entropy, which is additive ~\cite{GellMannT04}.  The Boltzmann factor is generalized into a {\it power-law}.
The mathematical basis for Tsallis statistics includes $q$-generalized expressions for the logarithm and the exponential functions which are the $q$-logarithm and the $q$-exponential functions. The $q$-{\em exponential function}, which reduces to $exp(x)$ in the limit $q$ $\rightarrow 1$, is defined as follows
\begin{equation}\label{eq:tsallis q-exp}
    e_{q}^{x}\equiv [1+(1-q) x]^{\frac{1}{(1-q)}}=\frac{1}{[1-(q-1)x]^{\frac{1}{(q-1)}}}  \;\;\;\;(e_1^x=e^x)  \, .
\end{equation}
We remind that extremizing entropy $S_{q}$  under appropriate constraints we obtain a probability distribution, which is proportional to $q$-{\em exponential function}.

In this work, we focus on the analysis of the distribution of citations of scientific publication, more precisely those that have been catalogued by the Institute for Scientific Information (ISI).
In [2000] Tsallis and Albuquerque \cite{TsallisMA00} suggested that the citation phenomenon appears to be deeply related to thermostatistical nonextensivity.
However, they conclude that it is important to understand what physical mechanism of the nonlinear dynamics of this phenomenon is responsible
for the specific values of $q$, which fit the experimental data. In their discussion they tried to understand
why a stretched exponential form does not fit the entire experimental range when citations per paper were focused,
whereas it appears to be satisfactory when citations per scientist were focused instead.
\subsection{Thomson ISI Web of Science}
Traditionally, the most commonly used source of bibliometric data is Thomson ISI Web of
Knowledge, in particular the (Social) Science Citation Index and the Journal Citation Reports
(JCR), which provide the yearly Journal Impact Factors (JIF) \cite{WOS}.
The subject categories and terminology provided by ISI are widely recognized by many researchers and
scientometricians in their studies and it is relatively simple to use \cite{BatistaCKM06,PetricekCHCG05,MaestroSC08}.

The Institute for Scientific
Information has made an industry of providing citation data to libraries
since the mid-1960s; the products are currently available as part of Thomson/ISI. ISI reports that they currently index more than 16,000 journals, books
and proceedings. All articles appearing in selected
publications have their bibliographies manually transcribed, and "inverted
bibliographies" pointing from a (earlier) cited work to all (subsequent) citing
publications is generated to support users' searches.
Various reports extensive analyse the scientific activity of particular countries in comparison with other countries, primarily based on ISI 's data \cite{PetricekCHCG05}. Although the ISI database
has a few shortcomings, overall it gives a wide coverage of most research fields \cite{Belew05,May97}. Therefore in our survey
we utilize Thomson/ISI Web of Science database to study the distribution of the citation within a variety of countries.

\section{Experimental Study}
\label{sec:WOS1}

The study of citation distributions helps to understand better the mechanics behind
citations and objectively compare scientific performance. In this section, the Tsallis \emph{q}-exponential distribution is used to describe the
Institute of Scientific Information citations for different countries around the world since 1945 to November 2008. More precisely, the distributions of citations related to $13$ different countries from Latin America, South Africa and Europe are exhibited and discussed. All of them have quite large data sets.

The search has been conducted using the query ``name of the tested country'' in the address field.
One of the reasons behind the specific choice of these countries for this work is the fact that the present version of
ISI Web of Science limits the searching to $100\,000$ papers annually, thus making it impossible to
compile a complete database for countries that have a greater annual productivity as the
United States of America, Japan, or France, etc. Some of the countries with an important number of publications,
such as Italy, Spain, Switzerland, Brazil and Austria, are chosen. Additionally, Greece, Hungary and Portugal from the European Union are selected as they have a quite large dataset and perform significant research activity and productivity. Romania is included in this survey as it is a new member of the European Union with
quite a few number of publications and has recently intensified research activity. Moreover, our database has data from Argentina, Chile and Mexico since they constitute countries of Latin America with noteworthy scientific performance. Finally, South Africa's research activity is investigated in order to include countries from different continents. Our database contains information of $3\,399\,572$
bibliographical references. This information includes the type of publication, full reference,
citations received, authors' names and addresses, including the institutions, cities, states
and countries.

In our fitting distribution papers with zero citations ($c=0$) were not taken into account \cite{PetricekCHCG05}. This is common as a significant number of new papers takes time to be on the database and needs time to start being cited. To mitigate against this distortion, we limit ourselves in both datasets to papers that have been cited at least once \cite{PetricekCHCG05}. Furthermore, the experimental study does not take primarily into consideration the papers with one citation to fit the data; nevertheless, its probability is taken into account for our final results (linear regression coefficient $R^2$).
The proposed fitting distributions follows from the nonextensive formalism as  $N(x) \propto 1/[1+(q-1)\; c/T]^{\frac{1}{q-1}}$.
In our experimental study we adopt the following expression:
\begin{equation}
N(c) = N(2)\, \exp_{q}^{-\frac{c-2}{T}}
\end{equation}
where $N(2)$ is the number of papers with two citations, and, as already mentioned, $T$ plays the role of an effective temperature.

All the tested countries are shown to have a satisfactory fitting with values of
$q$ in the following range: $q=1.34\pm 0.10$. Moreover the ``effective temperature'' values are defined within an error bar $\pm 0.05$. In order to take the optimal value of $q$ we use roughly the first two decades of citation data points for each country. In the next subsection, it is given detailed description of how the data collection is done, and all the results are presented for each country that is included in this survey.

\subsection{Data Acquisition and Treatment.}
\label{sec:Data Collection}
To obtain all the necessary data we developed a program which automatically downloads the ISI
bibliographic information.
We take into account all the document types, e.g. articles, proceeding papers, meeting abstracts, etc, for all the
available subject areas, for instance neurosciences, mathematics, chemistry
etc. The program was written in Delphi 7 and uses the TWebBrowser component.
This component provides access to the Web browser functionality and saves all
the ``html'' pages. When the page was completely downloaded, an
OnDownloadComplet event is generated and we go automatically to the next
``html'' page. When all pages were downloaded we process each ``html'' page to
obtain the specific information that we are interested in using the TPerlRegEx
component from the open source PCRE library [http://www.pcre.org/]. In this
case, we gather the number of citations for each publication and the total number of the published papers, for each
country. We apply filters to take all these data sorted by the times cited,
using the Citation databases namely Science Citation Index Expanded
($SCI-EXPANDED$ 1945--present), Social Sciences Citation Index ($SSCI$
1956--present, and Arts and humanities Citation Index ($AandHCI$
1975--present). All the data were captured during November 2008.

\subsection{Fitting the citations}
\label{sec:All Countries}
\begin{figure}
\includegraphics[width=0.95\textwidth]{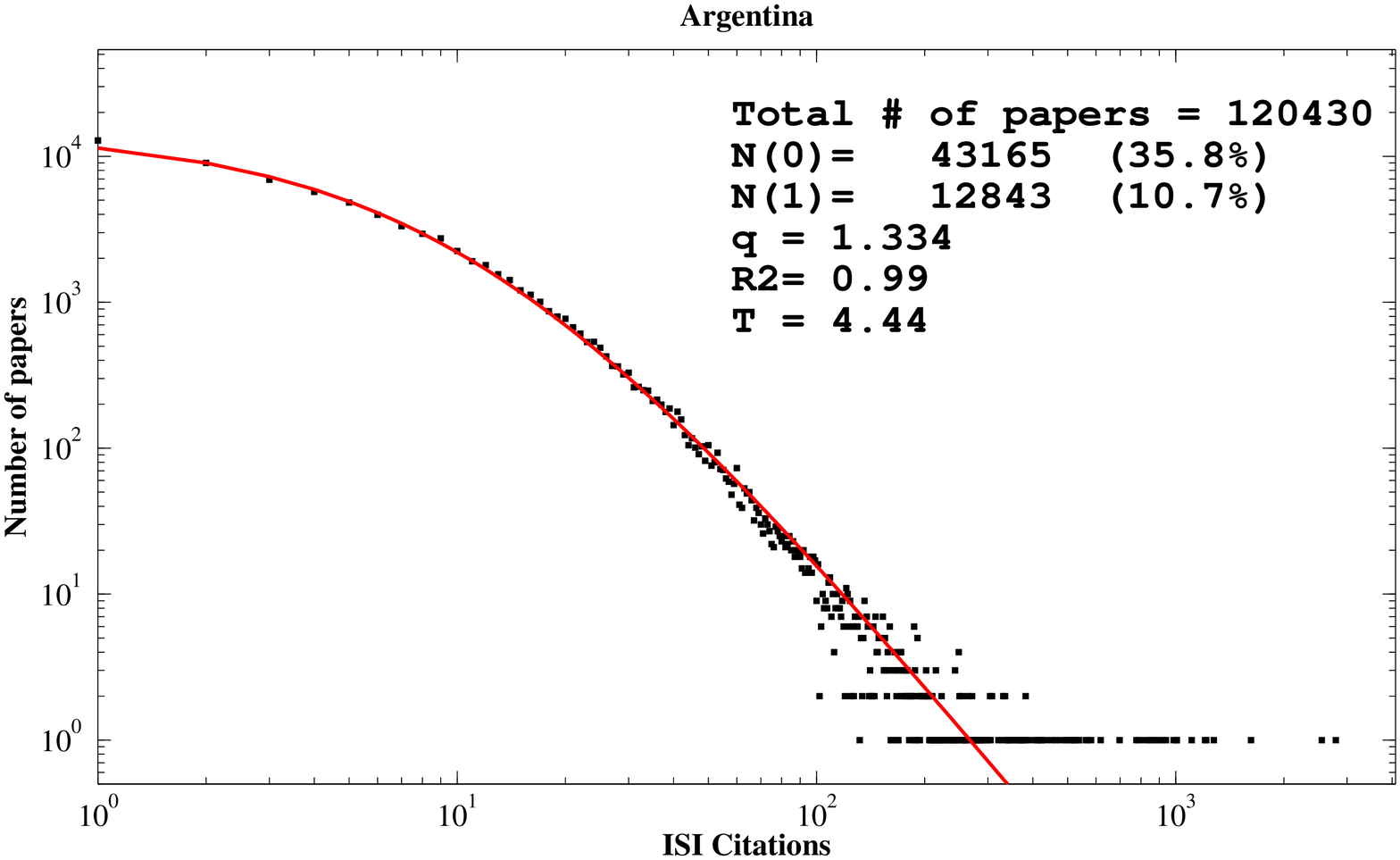}\\
\includegraphics[width=0.95\textwidth]{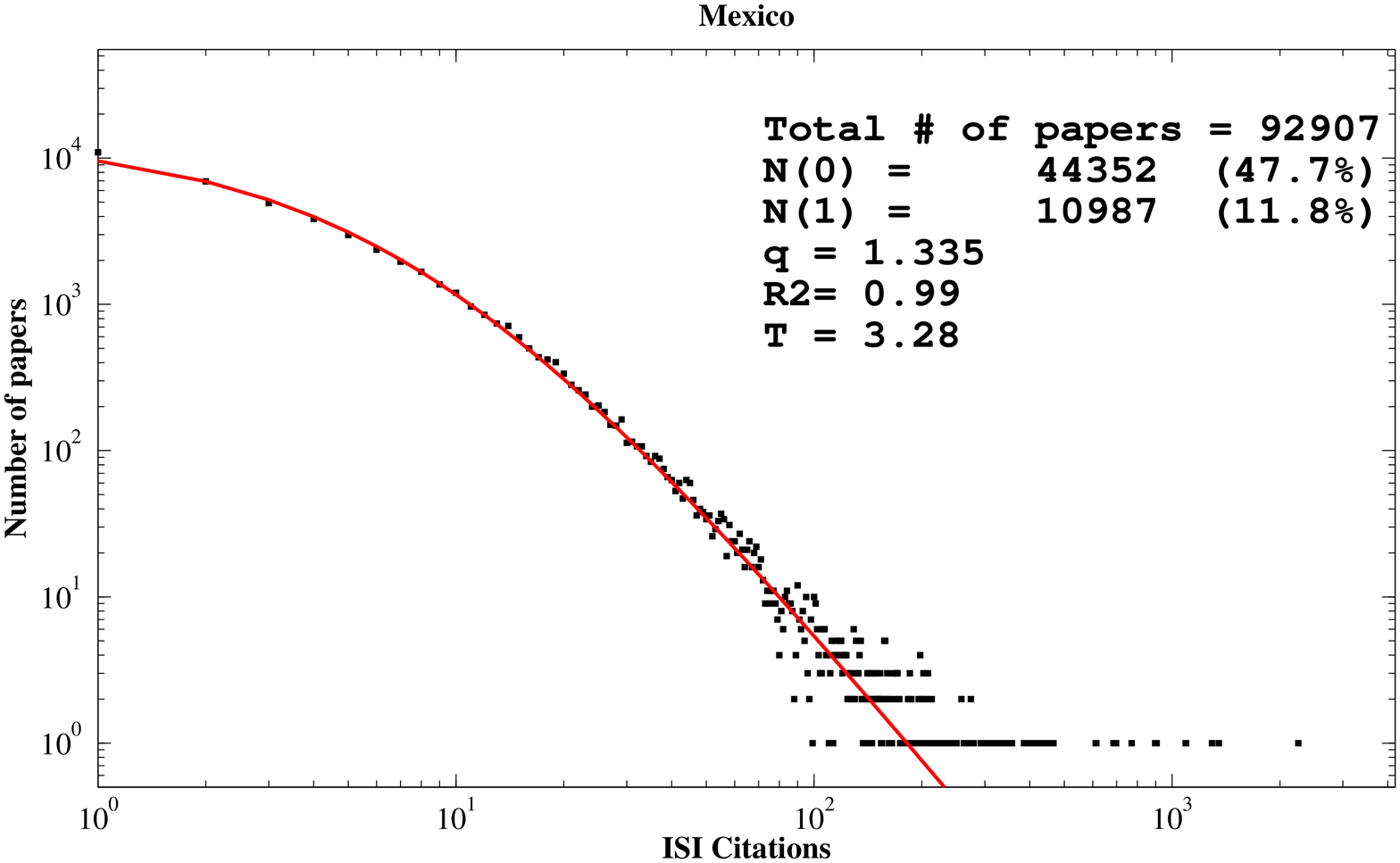}
\caption{Probability distribution for citations of Argentina (up) and Mexico (down) up to November 2008}
\label{fig:Mexico-Argentina}       
\end{figure}

Brazil was the first country within Latin America that we studied because of the high number of published papers \cite{TsallisBook08}. As we can observe in Table \ref{table:Total Papers}, the Brazilian total number of papers are $285\,570$. The same procedure that we followed for Brazil was applied for all the countries that are included in this work. For instance, in the case of Brazil, all papers included in the Web of Science and having at least one author with at least one affiliation address in Brazil have been collected. This means that the work includes all the documents with at least one Brazilian address with citations till November 2008.
Research done by Brazilians abroad, i.e with only foreign addresses, is disregarded in the considered database. Note that the data and results are presented on a log-log scale.
Initially we evaluate the values of $q$ in order to find its optimal value, and then, with this value, we do the final fitting in order to determine $T$.
With these two values ($q$ and $T$) we present the fitting in a log-log diagram. In the Brazil case a remarkably good fitting can be done with  $q=1.339$ and $T=3.97$. This temperature provides good evidence about the impact of the published papers, and enables a ranking. Figure \ref{fig:Mexico-Argentina} illustrates the fitting of two Latin American countries. Argentina achieves the highest temperature and Mexico the lowest among the tested Latin American countries. In both cases the nonextensive fitting distribution fits well the data. The \emph{q} value is almost the same giving temperature $T=4.44$ and $T=3.27$ for Argentina and Mexico respectively. It is important to mention that the $N(c)$ distribution is fitting well the data for all the Latin America countries and also provides an overall temperature. As we can notice from the Table \ref{table:Ranking Countries}, Brazil, because of its large amount of publications, has more effect than the other Latin American countries on the
temperature.

Table~\ref{table:Total Papers} shows the total number of the published papers, and the number of the publications with zero, one and two citations.
It also presents the countries ordered from the highest total number of publications (Italy) to the lowest total number of publications (Chile).
Furthermore, we present an overall outline for Latin American and European countries. A total number of 2\,677\,381 papers are published from European countries. Italy, Spain and Switzerland have the lowest percentage of the total number of papers with zero, one and two cited papers while Romania has 52.1\% of zero cited publications. Moreover, the typical number of the zero cited papers is around 35.0\% of the total number of publications for each country, while the usual amount of the one and two cited publications is about 10.5\% and 7.5\% respectively.
Finally, from Table~\ref{table:Total Papers}, it can be noticed that the average percentage of the zero, one and two cited publications for the
European tested countries is significantly less compared to Latin American tested countries.
\begin{table}
\vspace*{0.05cm}
\begin{center}
\begin{scriptsize}
\caption{Number of total publications, and the percentage of zero, one and two cited papers for the tested countries} \label{table:Total Papers}
\newcommand{\m}{\hphantom{$-$}}
\newcommand{\cc}[1]{\multicolumn{1}{c}{#1}}
\renewcommand{\tabcolsep}{0.7pc} 
\renewcommand{\arraystretch}{1.0} 
\begin{tabular}{@{}lllllllll}
\hline
            &  \textbf{Total \# Papers}     & $\;\;$\textbf{\# Zero citations}        &\textbf{\# One citations}     & \textbf{\# Two citations}   \\
Country     & $\;$$\sum_{c=0}^{\infty} N(c)$    & $\;\;$$N(0)$ $\;\;\;\;\;\;\;\;$ $(\%)$   &$\;\;$$N(1)$ $\;\;\;\;$ $\;\;$ $(\%)$  & $\;\;$$N(2)$ $\;\;\;\;$$( \% )$  \\
\hline
Italy                       &$\;\;$935\,769     &$\;\;\;$279\,013     $\;\;\;$(29.8\%)  &$\;\;$88\,477   $\;\;\;$($\;$9.5\%) &$\;\;\;$62\,543   (6.7\%)    \\
Spain                       &$\;\;$577\,996     &$\;\;\;$171\,696     $\;\;\;$(29.7\%)  &$\;\;$60\,530   $\;\;$(10.5\%)      &$\;\;\;$42\,205   (7.3\%)    \\
Switzerland                 &$\;\;$479\,642     &$\;\;\;$145\,546     $\;\;$  (30.3\%)  &$\;\;$41\,959   $\;\;$(8.70\%)      &$\;\;\;$27\,931   (5.8\%)     \\
Brazil                      &$\;\;$285\,570     &$\;\;\;$108\,984     $\;\;$ (38.2\%)   &$\;\;$33\,428   $\;\;$(11.7\%)      &$\;\;\;$22\,463   (7.9\%)       \\
Austria                     &$\;\;$237\,032     &$\;\;\;\;\;$83\,749  $\;\;\;$(35.3\%)  &$\;\;$24\,312   $\;\;$(10.3\%)      &$\;\;\;$16\,032   (6.8\%)       \\
South Africa                &$\;\;$157\,397     &$\;\;\;\;\;$55\,836  $\;\;\;$(35.5\%)  &$\;\;$17\,818   $\;\;$(11.3\%)      &$\;\;\;$12\,066   (7.7\%)     \\
Hungary                     &$\;\;$152\,385     &$\;\;\;\;\;$57\,209  $\;\;\;$(37.5\%)  &$\;\;$17\,087   $\;\;$(11.2\%)      &$\;\;\;$11\,251   (7.4\%)       \\
Greece                      &$\;\;$144\,443     &$\;\;\;\;\;$50\,235  $\;\;\;$(34.8\%)  &$\;\;$15\,753   $\;\;$(10.9\%)      &$\;\;\;$11\,166   (7.7\%)         \\
Argentina                   &$\;\;$120\,430     &$\;\;\;\;\;$43\,165  $\;\;\;$(35.8\%)  &$\;\;$12\,843   $\;\;$(10.7\%)      &$\;\;\;\;$9\,013  (7.5\%)  \\
Mexico                      &$\;\;\;\;$92\,907  &$\;\;\;\;\;$44\,352  $\;\;\;$(47.7\%)  &$\;\;$10\,987   $\;\;$(11.8\%)      &$\;\;\;\;$5\,189  (7.5\%) \\
Portugal                    &$\;\;\;\;$79\,988  &$\;\;\;\;\;$26\,010  $\;\;\;$(32.5\%)  &$\;\;\;$8\,383  $\;\;\;$(10.5\%)    &$\;\;\;\;$5\,841  (7.3\%) \\
Romania                     &$\;\;\;\;$70\,126  &$\;\;\;\;\;$36\,552  $\;\;\;$(52.1\%)  &$\;\;\;$8\,724  $\;\;\;$(12.4\%)    &$\;\;\;\;$5\,189  (7.4\%)   \\
Chile                       &$\;\;\;\;$65\,886 &$\;\;\;\;\;$25\,345   $\;\;\;$(38.5\%)  &$\;\;\;$7\,183  $\;\;\;$(10.9\%)    &$\;\;\;\;$4\,638  (7.0\%) \\
Latin American              &$\;\;$564\,794    &$\;\;\;$221\,847      $\;\;\;$(39.3\%)  &$\;\;$64\,441    $\;\;$(11.4\%)     &$\;\;$43\,045     (7.6\%) \\
European                    & 2\,677\,381      &$\;\;\;$849\,779      $\;\;\;$(31.7\%)  &265\,256       $\;\;$(9.90\%)       & 182\,161         (6.8\%)  \\
All Countries               & 3\,399\,572      &1\,127\,462           $\;\;\;$(33.2\%)  &347\,515       $\;\;$(10.2\%)       & 237\,272         (7.0\%)        \\
\hline
\end{tabular}
\end{scriptsize}
\end{center}
\vspace*{0.0cm}
\end{table}

\begin{figure*}
\includegraphics[width=0.95\textwidth]{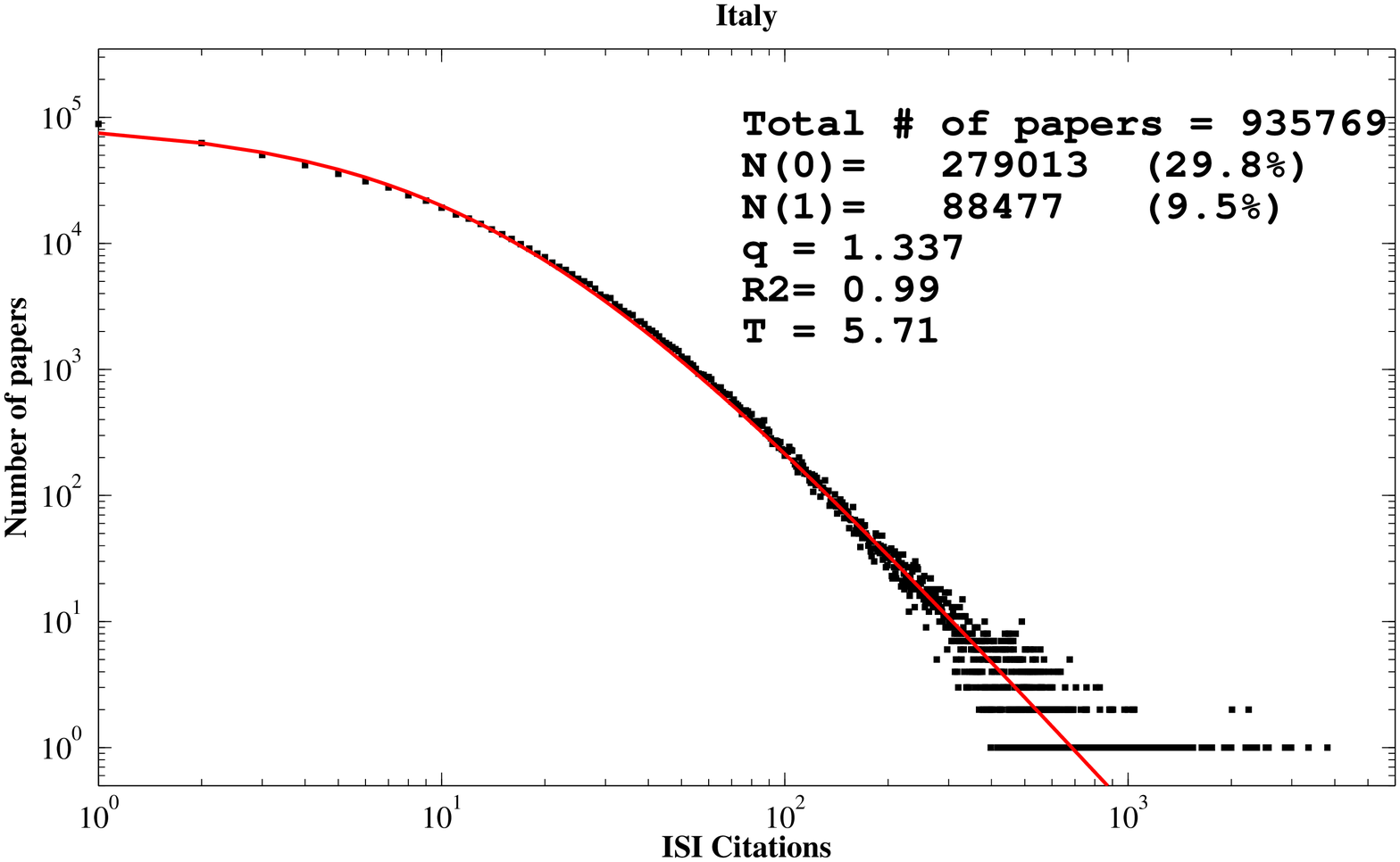}\\
\includegraphics[width=0.95\textwidth]{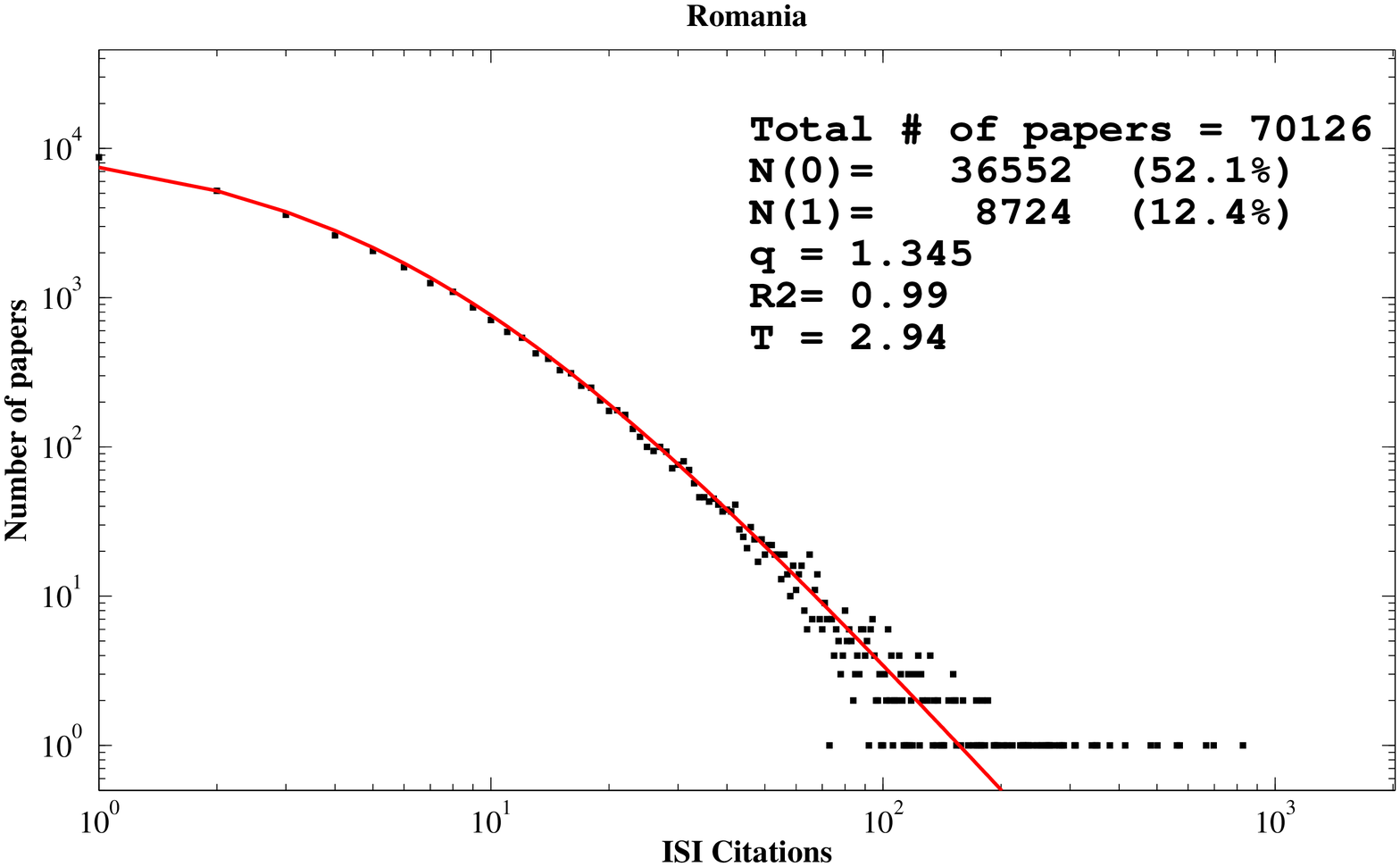}
\caption{Probability distribution for citations of Italy (up) and Romania (down) up to November 2008}
\label{fig:Italy-Romania}       
\end{figure*}

Table \ref{table:Ranking Countries} presents the countries in the ranking based on the temperature that we obtain through the nonextensive distribution fitting.
Notice that this ranking differs from the one presented on Table ~\ref{table:Total Papers}, where the total amount of the published papers (quantity ranking) is shown. The effective temperature $T$ characterizes the scientific impact of the tested countries. As we can perceive from Table \ref{table:Ranking Countries}, in all cases the range value of the entropic index \emph{q} is around $q=4/3$.
The linear regression coefficient $R^2$ is also indicated in each case.

As we can see comparing Tables 1 and 2, the rankings are quite different. Let us check Chile, for instance. Although it has a relatively small number of papers (Table ~\ref{table:Total Papers}, 65\,886 papers), its effective temperature is high $T=4.35$.

On the other hand, Italy and Spain exhibit a high ranking for both the impact and the quantity of their publications. Argentina attains the uppermost temperature in Latin America. Romania which has recently become a member of the European Union has the lowest values for the quantity of published papers and effective temperature. In this sense, Greece, Hungary, Portugal and South Africa achieve intermediate effective temperatures ($T=4.41, 4.40, 4.65, 4.25$ respectively).
Finally, Austria has an effective temperature close to the highest temperatures among all the tested countries.

\begin{figure*}
\includegraphics[width=0.95\textwidth]{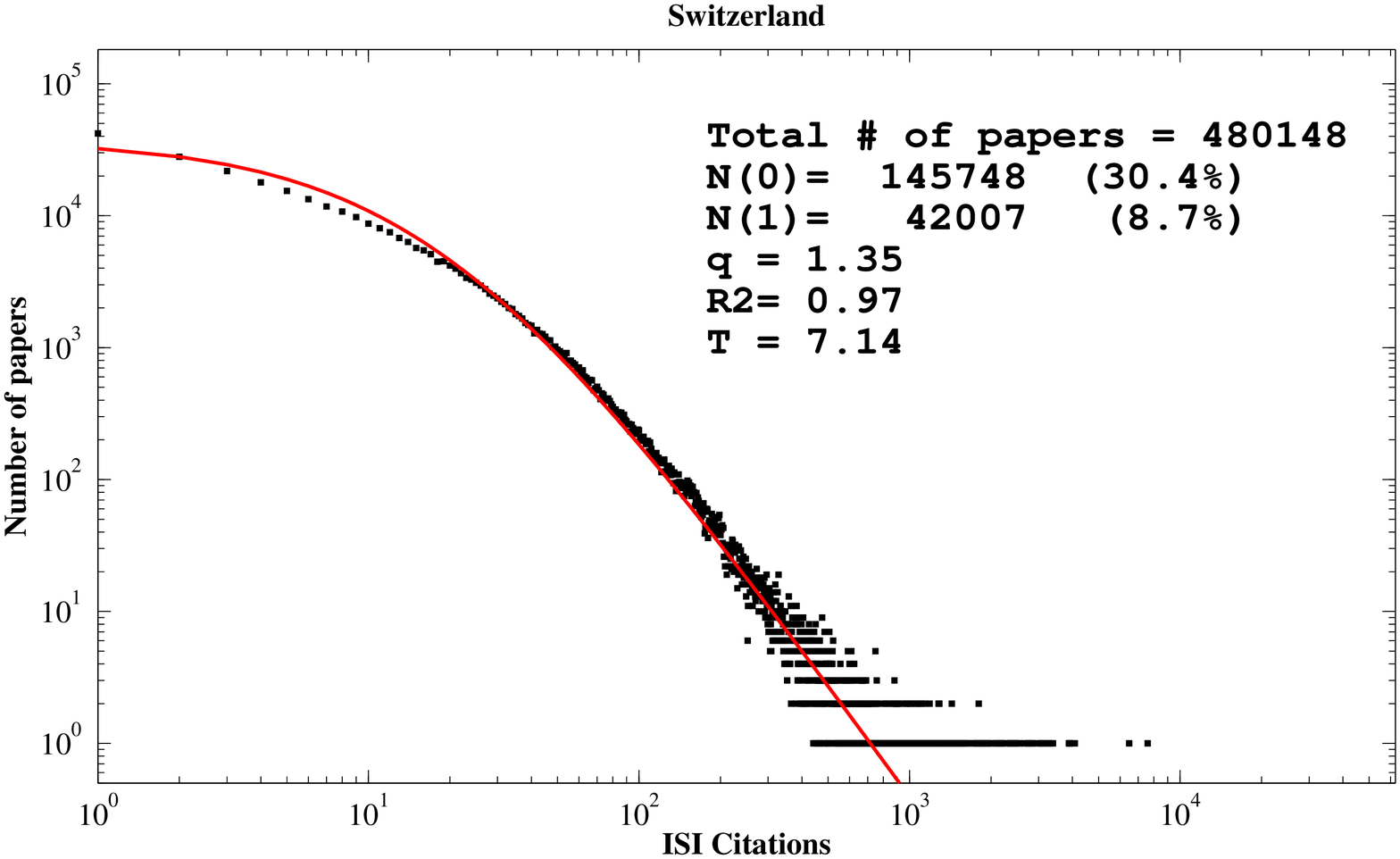}\\
\includegraphics[width=0.95\textwidth]{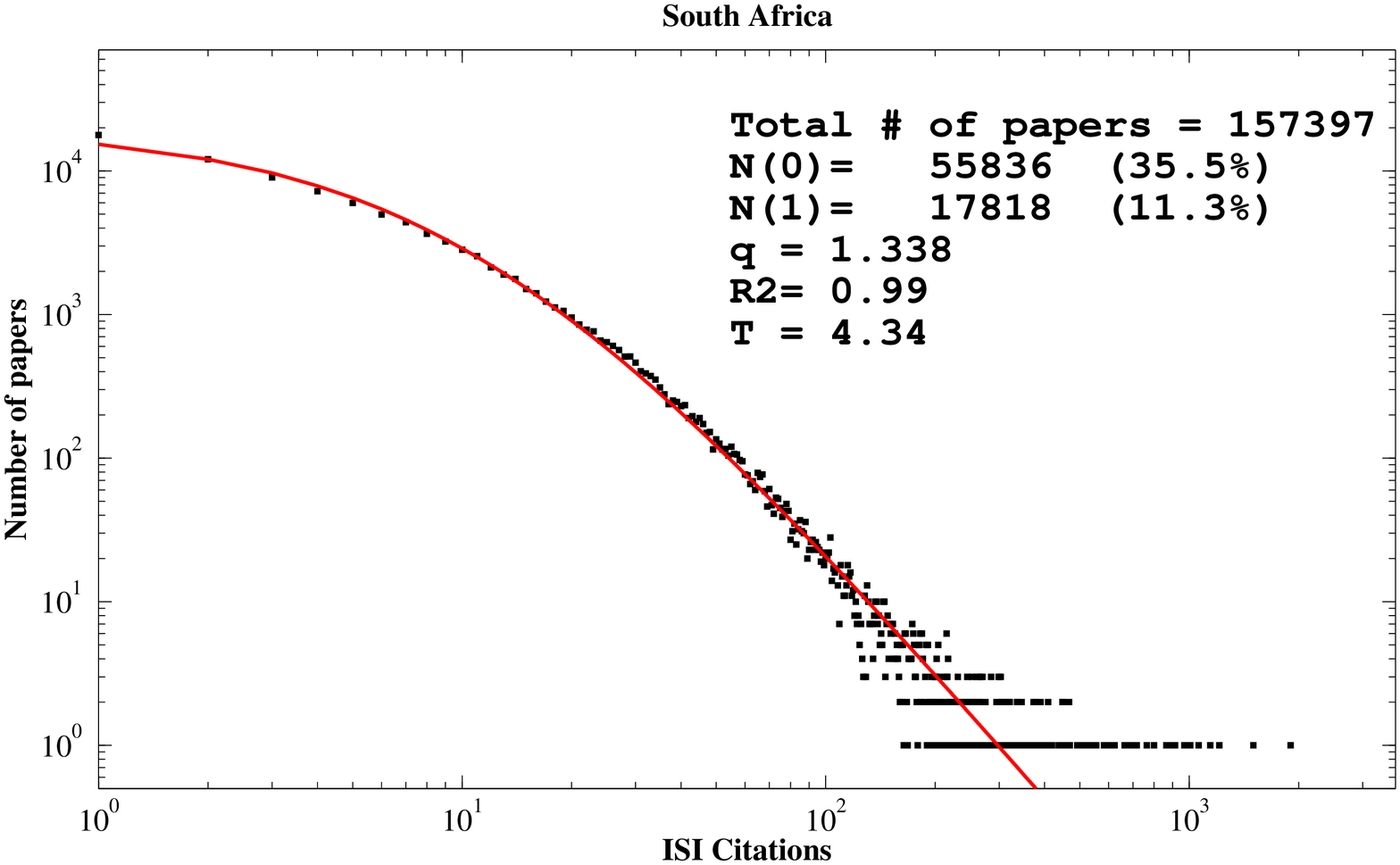}
\caption{Probability distribution for citations of Switzerland (up) and South Africa (down) up to November 2008}
\label{fig:Switzerland-South Africa}       
\end{figure*}

At this point it is important to mention the overall performance of Switzerland.
As we may notice from the Table \ref{table:Ranking Countries}, Switzerland achieves the highest effective temperature (with $R^2=0.97$); see
Figure \ref{fig:Switzerland-South Africa}.
\begin{table*}
\vspace*{0.05cm}
\begin{center}
\begin{small}
\caption{Best fitting values of \emph{q} and effective temperature\emph{T}. Note that tested countries are ranked according to \emph{T}} \label{table:Ranking Countries}
\newcommand{\m}{\hphantom{$-$}}
\newcommand{\cc}[1]{\multicolumn{1}{c}{#1}}
\renewcommand{\tabcolsep}{0.6pc} 
\renewcommand{\arraystretch}{1.0} 
\begin{tabular}{@{}lllllllllll}
\hline
                &   \textbf{Entropic index}           & \textbf{Linear regression}    & \textbf{Temperature}    \\
Country         & $q$                 & {\bf coefficient} $R^{2}$                      & \textbf{$T$}           \\
\hline
Switzerland                 & 1.350                             & 0.97                      & \textbf{7.14}            \\
Italy                       & 1.337                             & 0.99                      & \textbf{5.82}             \\
Spain                       & 1.325                             & 0.99                      & \textbf{5.20}              \\
Austria                     & 1.400                             & 0.99                      & \textbf{4.87}             \\
Portugal                    & 1.336                             & 0.99                      & \textbf{4.65}             \\
Argentina                   & 1.334                             & 0.99                      & \textbf{4.44}            \\
Greece                      & 1.330                             & 0.99                      & \textbf{4.41}           \\
Hungary                     & 1.339                             & 0.99                      & \textbf{4.40}                \\
Chile                       & 1.350                             & 0.98                      & \textbf{4.35}            \\
South Africa                & 1.338                             & 0.99                      & \textbf{4.25}            \\
Brazil                      & 1.343                             & 0.99                      & \textbf{3.97}              \\
Mexico                      & 1.335                             & 0.99                      & \textbf{3.28}            \\
Romania                     & 1.345                             & 0.99                      & \textbf{2.94}              \\
Latinamerican countries (4) & 1.330                             & 0.99                      & \textbf{4.08}              \\
European countries (8)      & 1.334                             & 0.99                      & \textbf{5.88}            \\
All countries (13)          & 1.340                             & 0.99                      & \textbf{5.26}             \\
\hline
\end{tabular}
\end{small}
\end{center}
\vspace*{0.0cm}
\end{table*}
We can observe in the figure \ref{fig:Switzerland-South Africa} that, in contrast with the other countries, the nonextensive distribution
$N(c)$ can not fit accurately the first data points (total number of publications with two and three citations); for the rest of the points, we get satisfactory
fitting. The curve is crossing slightly above of these points. In order to study this issue further we analyze two distinguished institutions, namely
the Harvard University and Oxford University. Both of these universities have high effective temperatures (7.92 and 7.69 respectively). Furthermore, their plots are similar to those of Switzerland. Summarizing, in the cases of extremely high impact research activity, there is a common behavior, reflecting in the first data points.

In figure \ref{fig:Switzerland-South Africa} South Africa is also presented. In this case the nonextensive distribution fits well with \emph{q=1.338} and an effective temperature  $T=4.25$. Figure \ref{fig:Italy-Romania} shows the graphs for Italy and Romania.
Although there is a significant difference in the amount of published papers and in the effective temperature between these two countries, the $N(c)$ distribution achieves a fine matching of their data points.

From all the above experimental results, we obtain a value of $q$ close to $4/3$. For all these values of $\textit{q}$ there is no significant effect on the temperature. For instance, if we apply for Italy $\emph{q=1.40} $ or $\emph{q=1.330}$  instead of $\emph{q=1.337}$ we obtain essentially the same effective temperature (with only a tiny change in $R^2$, 0.991 instead of 0.994).
\begin{figure*}[htp]
\includegraphics[width=0.95\textwidth]{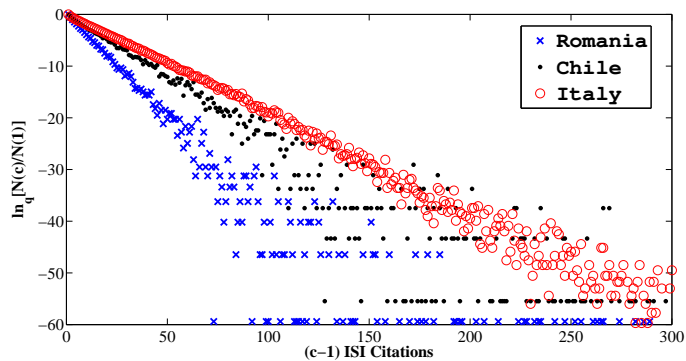}
\caption{$\ln_{q}[{N(c)/N(1)}]$ versus $(c-1)$ for 3 countries (Romania, Chile, Italy) from 1945 to November 2008. 
The figure is a zoom with an upper limit of 300 citations.}
\label{fig:Linear presentation tested Countries}       
\end{figure*}
\begin{figure*}[htp]
\includegraphics[width=0.95\textwidth]{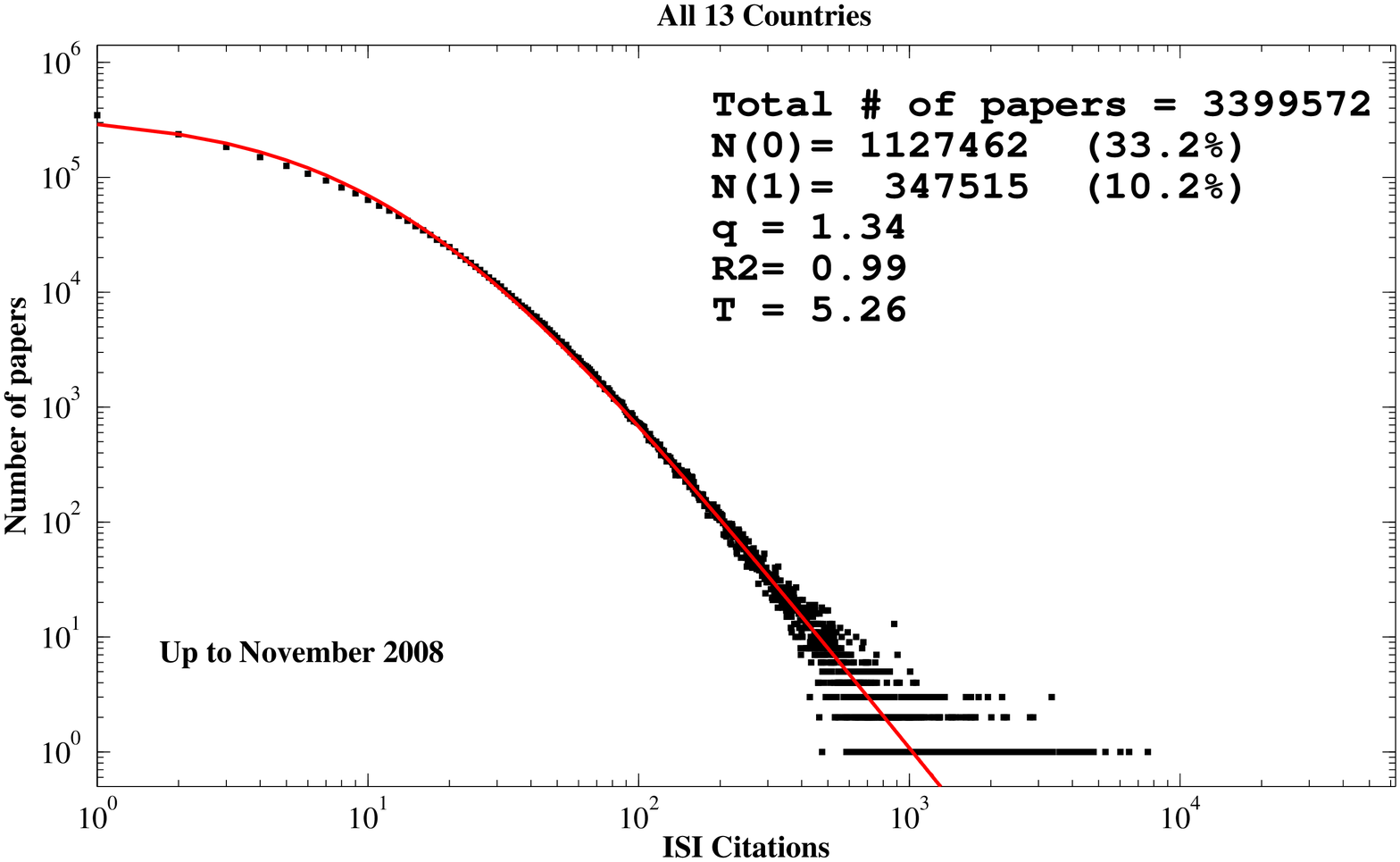}
\caption{Probability distribution for citations of all 13 tested countries from 1945 up to November 2008}
\label{fig:All 13 tested Countries}       
\end{figure*}

Finally, Figure \ref{fig:Linear presentation tested Countries} illustrates the $q$-logarithmic number of publications $\ln_{q}[{N(c)/N(1)}]$  versus the $(c-1)$ number of citations for three countries (Romania, Chile, Italy). It is important to notice that $(-1/T)$ corresponds to the average slope associated with each country. The best fitting values of $q$ (see Table \ref{table:Ranking Countries}) were used for the plotting of $\ln_{q}[{N(c)/N(1)}]$ versus the $(c-1)$ for these three countries. Romania, Chile and Italy were selected because Romania achieves the lowest ``effective temperature'' $T$, Chile an intermediate  $T$, and Italy one of the highest in our experimental study. Through this selection we can clearly observe the behavior of the data by just using any value of $q$ close to 4/3.
As you can see from the Figure \ref{fig:Linear presentation tested Countries} the different slopes associated with different countries follow monotonically our proposed impact ranking ($T$) (in this case, the temperature for Romania is $2.96$, for Chile $4.35$, and for Italy $5.82$).

\section{Concluding Remarks and Discussion}
\label{Remarks}
Nowadays the number of citations is among the most widely used measures of academic performance. Extended study of citation distributions helps to understand better the mechanics behind citations and objectively can establish a comparative measure for scientific performance. Citations of scientific papers is in fact a connection network constituted by authors (nodes) and directed links (citations) among them. Recently, connection networks have been described, studied, characterized and represented by parameters using typical concepts in the area of Complex Systems. Soares et al \cite{SoaresTMS05} studied a two-dimensional growth model to characterize a network where every new site is placed, at a distance $r$ from the barycenter of a pre-existing graph.
These networks depend on two basic principles: continuous growth of the number of nodes and preferential attachment connections. A preferential attachment connection is when nodes which already have a high number of connections (hubs) have bigger probability of receiving new connections.
The linking is done with a probability which decays with distance as $1/r^\alpha$ ($\alpha \ge 0$).
They showed that the microscopic dynamics of this nonextensive system typically builds an (asymptotically) scale-free network, and when $\alpha \to 0$, the entropic index $q$ precisely approaches $4/3$, thus recovering the well known Barabasi-Albert class.

The entropic index $q$ in Tsallis entropy is usually interpreted as a quantity characterizing the degree of nonextensivity of a system. An appropriate choice of the entropic index $q$ to nonextensive physical systems still remains an open field of study.
In some cases, the physical meaning of the index $q$ is unknown; it provides nevertheless
new possibilities of comparison between theoretical approaches and experimental data. Other cases are better understood, and then $q$ has a clear
physical meaning, either at a microscopic or at a mesoscopic level, or both.

In this paper we propose the Tsallis $q$-exponential distribution to satisfactorily describe Institute of Scientific Information citations between 1945 and November 2008. Our study provides evidence that the citation distribution for all tested cases within this period could be the Tsallis $q$-exponential distribution. Figure \ref{fig:Linear presentation tested Countries} gives an explanation for the meaning of $T$, and the ranking that we proposed based on the $T$. Finally, this manuscript provides a new ranking index for the impact level of the research activity for the $13$ tested countries via the ``effective temperature'' $T$, a new characterization of impact.

Figure \ref{fig:All 13 tested Countries} exemplifies the overall number of citations for all the tested countries in this survey. It is remarkable how the proposed nonextensive distribution fits satisfactorily all the amount of the cited papers for all the countries. This part of the analysis uses the entire available-year publication window for all disciplines for papers published between 1945 to November 2008. Future work can address  how the scientific research activity changes with time, for instance between 1945 to 1990, then during the period 1991-2000, and finally from 2001 to nowadays.

\begin{acknowledgements}
The authors thank Professor C.Tsallis, S. Thurner, R. Hanel and D. Kalamatianos for helpful discussions and the support from the National Council for Scientific and Technological Development (CNPq) of the Brazilian Ministry of Science and Technology,
the State of Rio de Janeiro Research Foundation (FAPERJ) and the Brazilian Coordination Office for the Improvement of Staff with Higher Education (CAPES).
\end{acknowledgements}



\end{document}